\documentclass[12pt]{article}
\usepackage{jheppub}

\pdfoutput=1

\usepackage{amsmath,bbm,array,amsfonts,graphicx,wrapfig,lscape,float,mathtools,multirow,longtable}
\usepackage[dvipsnames]{xcolor}

\newcommand{\be}{\begin{equation}}
\newcommand{\ee}{\end{equation}}
\newcommand{\beq}{\begin{equation}}
\newcommand{\beql}[1]{\begin{equation}\label{#1}}
\newcommand{\eeq}{\end{equation}}
\newcommand{\ba}{\begin{array}}
\newcommand{\ea}{\end{array}}
\newcommand{\bea}{\begin{eqnarray}}
\newcommand{\beal}[1]{\begin{eqnarray}\label{#1}}
\newcommand{\eea}{\end{eqnarray}}
\newcommand{\ben}{\begin{enumerate}}
\newcommand{\een}{\end{enumerate}}
\newcommand{\bean}{\begin{eqnarray*}}
\newcommand{\eean}{\end{eqnarray*}}

\newcommand{\tref}[1]{Table~\ref{#1}}

\newcommand{\fref}[1]{Figure \ref{#1}}
\newcommand{\btab}[1]{\begin{tabular}{#1}}
\newcommand{\etab}{\end{tabular}}

\def \Black {\pdfsetcolor{0 0 0 1}}

\newcommand{\comment}[1]{}

\newcommand{\qed}{\nobreak \ifvmode \relax \else
      \ifdim\lastskip<1.5em \hskip-\lastskip
      \hskip1.5em plus0em minus0.5em \fi \nobreak
      \vrule height0.75em width0.5em depth0.25em\fi}

\definecolor{darkspringgreen}{rgb}{0.09, 0.45, 0.27}
\definecolor{forestgreen}{rgb}{0.13, 0.55, 0.13}

\usepackage{array}
\usepackage{physics}
\usepackage{subcaption}
\usepackage{tikz,tikz-3dplot}
\usepackage[colorlinks=true]{hyperref}
\newcolumntype{C}[1]{>{\centering\let\newline\\\arraybackslash\hspace{0pt}}m{#1}}

\definecolor{yellow2}{rgb}{0.98, 0.80, 0.20}

\title{2d Supersymmetric Gauge Theories, D-branes and Trialities}

\author[a,b,c]{Sebasti\'an Franco}

\affiliation[a]{Physics Department, The City College of the CUNY\\
	160 Convent Avenue, New York, NY 10031, USA}
\affiliation[b]{Physics Program and \textsuperscript{$c$}Initiative for the Theoretical Sciences\\
	The Graduate School and University Center, The City University of New York\\
	365 Fifth Avenue, New York NY 10016, USA}
	
\emailAdd{sfranco@ccny.cuny.edu}

\abstract{Engineering quantum field theories in String Theory in terms of branes is a powerful approach for understanding their dynamics. We review recent progress in the realization of $2d$ $\mathcal{N}=(0,2)$ gauge theories in terms of branes. We discuss Brane Brick Models, a new class of Type IIA brane configurations which are T-dual to D1-branes over singular toric Calabi-Yau 4-folds. They fully encode the infinite class of $2d$ $\mathcal{N}=(0,2)$ quiver gauge theories on the worldvolume of the D1-branes and significantly streamline their connection to the probed geometries. As an application, we explain how these constructions provide a brane realization of triality. We also comment on the realization of $2d$ $\mathcal{N}=(0,1)$ theories via Spin(7) orientifolds. This note is based on the author's talk at the Nankai Symposium on Mathematical Dialogues celebrating the 110$^{th}$ anniversary of the birth of Prof. S.-S. Chern
}

\begin{document}

\maketitle

\section{Introduction} 

\begin{wrapfigure}{r}{0.275\textwidth}
  \vspace{-25pt}
  \begin{center}
    \includegraphics[width=2.5cm]{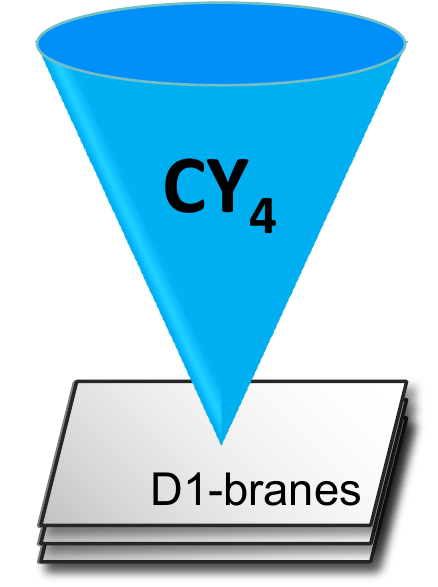}
  \end{center}
  \vspace{-20pt}
  \caption{D1-branes probing a CY$_4$.}
  \vspace{-10pt}
  \label{D1s_over_CY4}
\end{wrapfigure}
Understanding the dynamics of quantum field theories in various dimensions is an active area of research. For this purpose, engineering quantum field theories in terms of branes in String and M-Theory is often a powerful tool. In this note, we will review the realization of $2d$ $\mathcal{N}=(0,2)$ gauge theories on the world volume of D1-branes probing singular toric Calabi-Yau (CY) 4-folds \cite{Garcia-Compean:1998sla,Franco:2015tna}, as schematically shown in \fref{D1s_over_CY4}.

\section{Brane Brick Models}

\label{section_BBMs}


Brane brick models are obtained from D1-branes at $\text{CY}_4$ singularities by T-duality. We refer the reader to \cite{Franco:2015tna,Franco:2015tya,Franco:2016nwv,Franco:2016qxh} for detailed presentations. A brane brick model is a Type IIA brane configuration consisting of D4-branes wrapping a 3-torus $\mathbb{T}^3$ and suspended from an NS5-brane that wraps a holomorphic surface $\Sigma$ intersecting with $\mathbb{T}^3$. The holomorphic surface $\Sigma$ is the zero locus of the Newton polynomial defined by the toric diagram of the $\text{CY}_4$. The basic ingredients of the brane setup are summarized in Table \ref{Brane brick-config}. The $(246)$ directions are compactified on a $\mathbb{T}^3$. The $2d$ gauge theory lives on the two directions $(01)$ common to all the branes.

\begin{table}[ht!!]
\centering
\begin{tabular}{l|cccccccccc}
\; & 0 & 1 & 2 & 3 & 4 & 5 & 6 & 7 & 8 & 9 \\
\hline
$\text{D4}$ & $\times$ & $\times$ & $\times$ & $\cdot$ & $\times$ & $\cdot$ & $\times$ & $\cdot$ & $\cdot$ & $\cdot$  \\
$\text{NS5}$ & $\times$ & $\times$ & \multicolumn{6}{c}{----------- \ $\Sigma$ \ ------------} & $\cdot$ & $\cdot$ \\
\end{tabular}
\caption{Brane brick model configuration.}
\label{Brane brick-config}
\end{table}

Brane brick models, or equivalently their dual periodic quivers, fully encode the $2d$ $\mathcal{N}=(0,2)$ quiver gauge theories on the worldvolume of D1-branes probing toric CY 4-folds. Namely, they summarize not only the quivers but also the $J$- and $E$-terms. The dictionary between brane brick models and gauge theories is summarized in \tref{tbrick}. 

\begin{table}[H]
\centering
\resizebox{\hsize}{!}{
\begin{tabular}{|l|l|l|}
\hline
{\bf Brane Brick Model} \ \ &  {\bf Gauge Theory} \ \ \ \ \ \ \  & {\bf Periodic Quiver} \ \ \ 
\\
\hline\hline
Brick  & Gauge group & Node \\
\hline
Oriented face  & Bifundamental chiral field & Oriented (black) arrow 
\\
between bricks $i$ and $j$ & from node $i$ to node $j$  & from node $i$ to node $j$ \\
\hline
Unoriented square face  & Bifundamental Fermi field & Unoriented (red) line \\
between bricks $i$ and $j$ & between nodes $i$ and $j$ & between nodes $i$ and $j$  \\
\hline
Edge  & Interaction by $J$- or $E$-term & Plaquette encoding \\ 
& & a $J$- or an $E$-term \\
\hline
\end{tabular}
}
\caption{
Dictionary between brane brick models and $2d$ $\mathcal{N}=(0,2)$ gauge theories.
\label{tbrick}
}
\end{table}

Brane brick models reduce the computation of the underlying CY$_4$ geometry  starting from the gauge theory to a combinatorial problem, which is based on a generalization of perfect matchings \cite{Franco:2015tya}. Conversely, several efficient algorithms for determining the brane brick models associated to a given geometry have been developed \cite{Franco:2015tna,Franco:2016qxh,Franco:2016fxm,Franco:2018qsc,Franco:2020avj}.

\section{Triality}

\begin{figure}[ht]
	\centering
	\includegraphics[width=12cm]{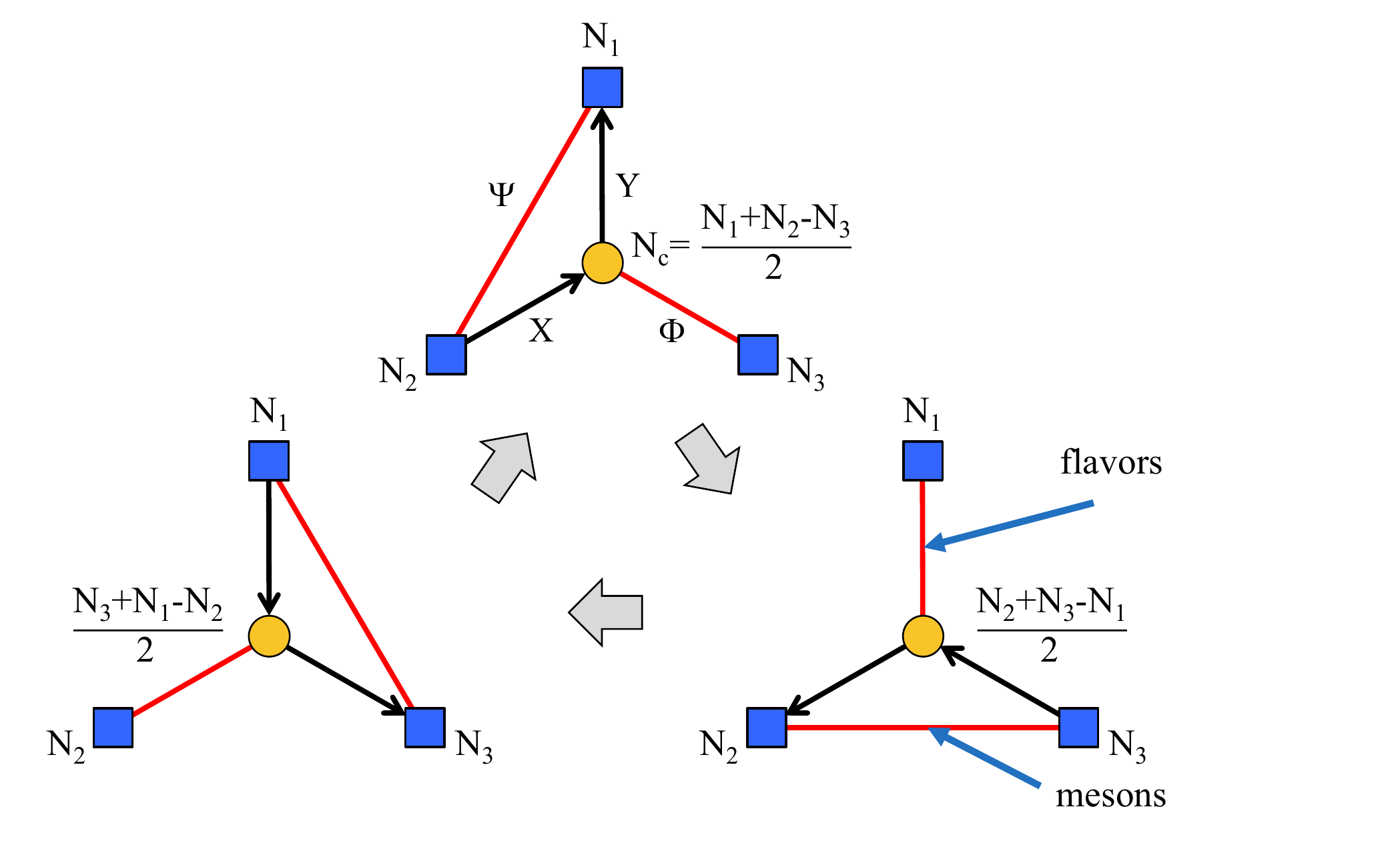}
\caption{Triality of $2d$  $\mathcal{N}=(0,2)$ SQCD. Black arrows and red lines represent chiral and Fermi fields, respectively.} 
	\label{Triality}
\end{figure}

A new IR equivalence for $2d$  $\mathcal{N}=(0,2)$ gauge theory was discovered in \cite{Gadde:2013lxa}. In its simplest incarnation, which is shown in \fref{Triality}, it relates three SQCD-like theories and was therefore dubbed {\it triality}.\footnote{In the figure, yellow nodes indicate $SU(N_c)$ gauge groups. The theories on D1-branes have instead $U(N_c)$ gauge groups. A $U(N_c)$ version of triality was also introduced in \cite{Gadde:2013lxa}. It differs from the $SU(N_c)$ triality in \fref{Triality} by the presence of additional Fermi fields in the determinant representation of the gauge group for the cancellation of the Abelian anomaly. It is expected that Abelian anomalies of gauge theories on D1-branes are cancelled via a generalized Green-Schwarz mechanism. For this reason, the determinant Fermi fields are not present in those theories and triality reduces to the one considered here.} In more general theories, triality can be locally applied to an individual gauge group, with the rest of the theory acting as a spectator. Applying triality to the same gauge group three consecutive times results in the original theory.

\begin{figure}[ht]
	\centering
	\hspace{3.5cm}\includegraphics[width=10cm]{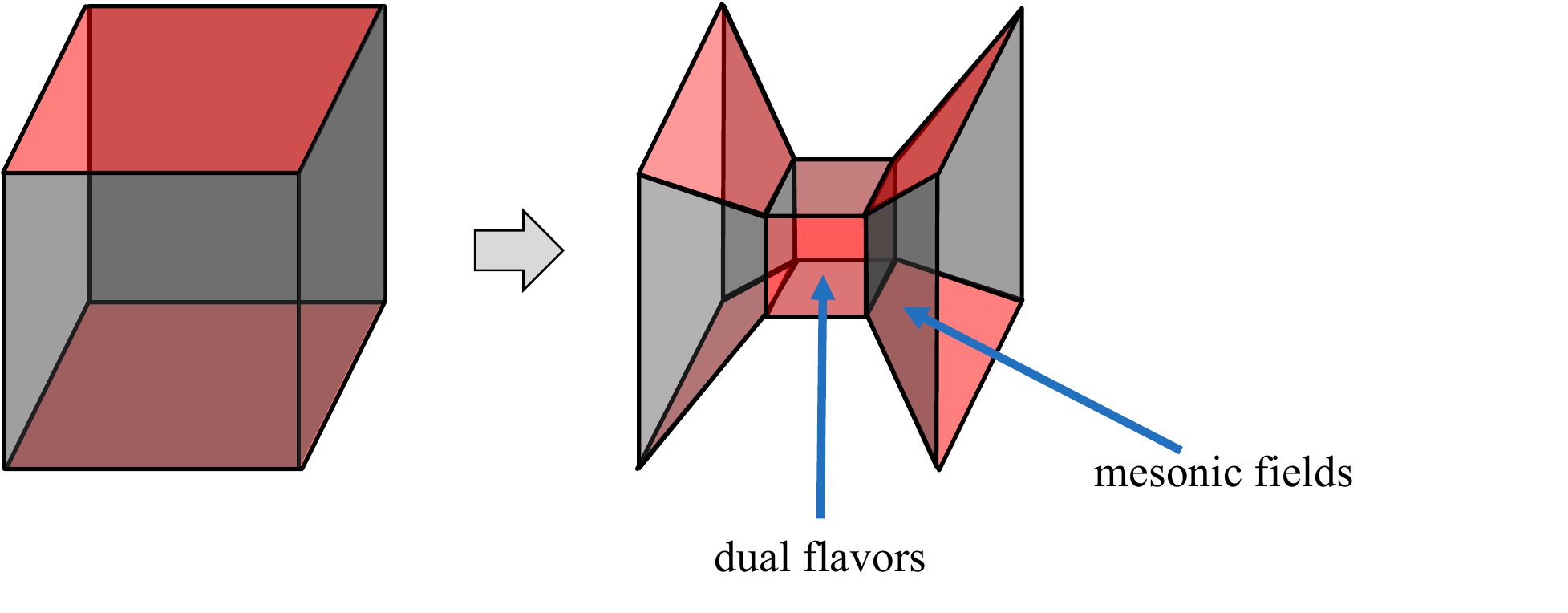}
\caption{The cube move is a local transformation of a brane brick models realizing $\mathcal{N}=(0,2)$ triality}
	\label{cube_move}
\end{figure}

For brane brick models, triality is elegantly realized by the local transformation shown in \fref{cube_move}, which we denote {\it cube move}. More precisely, the cube move is the simplest representative of a family of local transformations implementing triality \cite{Franco:2016nwv}. All these moves leave the underlying toric CY$_4$ invariant.

The correspondence between D1-brane probing toric CY 4-folds and brane brick models can be understood systematically using mirror symmetry \cite{Futaki:2014mpa,Franco:2016tcm}. The mirror configuration consists of D5-branes wrapping 4-spheres and the gauge theory is determined by how they intersect. In this context, triality is realized in terms of geometric transitions in the mirror geometry \cite{Franco:2016tcm}. More generally, mirror symmetry leads to a beautiful geometric unification of dualities in different dimensions, where the order of the duality is $n-1$ for a CY $n$-fold \cite{Franco:2016tcm,Franco:2017lpa}.

\section{Spin(7) Orientifolds}

As usual, it is desirable to investigate theories with less supersymmetry. The next step corresponds to $2d$ $\mathcal{N} = (0, 1)$, namely minimally supersymmetric, theories. These theories are extremely interesting, since they are barely supersymmetric and live at the borderline between non-SUSY theories and others with higher amounts of SUSY, for which powerful tools such as holomorphy become applicable. Due to the reduced SUSY, they enjoy a broad range of interesting dynamics. While there has been recent progress in their understanding, they remain relatively unexplored.

The geometric engineering of $2d$ $\mathcal{N} = (0, 1)$ gauge theories on D1-branes probing singularities was initiated in \cite{Franco:2021ixh}, where a new class of backgrounds, denoted {\it Spin(7) orientifolds} was introduced. These orientifolds are quotients of CY 4-folds by a combination of an anti-holomorphic involution leading to a Spin(7) cone and worldsheet parity. They naturally connect the view of $\mathcal{N} = (0, 1)$ theories as real slices of $\mathcal{N} = (0, 2)$ theories and Joyce’s geometric construction of Spin(7) manifolds starting from CY 4-folds \cite{Joyce:1999nk}. This geometric construction provides a new approach for studying $2d$ $\mathcal{N} = (0, 1)$ theories. It also gives a new perspective on the recently proposed triality of $2d$ $\mathcal{N} = (0, 1)$ gauge theories \cite{Gukov:2019lzi}, in which triality translates into the fact that multiple gauge theories correspond to the same underlying orientifold \cite{Franco:2021vxq}.

\acknowledgments

I would like to thank the organizers of the Nankai Symposium on Mathematical Dialogues for putting together such an exciting meeting and for the opportunity to present my work. I am indebted to my wonderful colleagues, Cyril Closset, Dongwook Ghim, Jirui Guo, Azeem Hasan, Sangmin Lee, Alessandro Mininno, Gregg Musiker, Rak-Kyeong Song, Angel Uranga, Cumrun Vafa, Daisuke Yokoyama and Xingyang Yu, for enjoyable collaborations on the topics presented in this note. This research was supported by the U.S. National Science Foundation grants PHY-1820721, PHY-2112729 and DMS-1854179.


\bibliographystyle{JHEP}
\bibliography{mybib}

\end{document}